\documentstyle[preprint,aps]{revtex}
\def\intpv{\mathop{\int{\mkern-17.5mu}-}}

\newcommand{\nablabf}{\mbox{\boldmath$\nabla$}}
\begin{document} 
\draft 
\title{On slip pulses at a sheared frictional viscoelastic/ non deformable 
interface}
\author{C. Caroli} 
\address{Groupe de Physique des Solides\cite{CNRS}, 2 place Jussieu, 
75251 Paris cedex 05, France.} 
\date{\today} 
\maketitle 

\begin{abstract} 
We study the possibility for a semi-infinite block of linear viscoelastic 
material, in homogeneous frictional contact with a non-deformable one, 
to slide under shear via a periodic set of ``self-healing pulses'', i.e. 
a set of drifting slip regions separated by stick ones. We show that, 
contrary to existing experimental indications, such a mode of 
frictional sliding is impossible for an interface obeying a simple 
local Coulomb law of solid friction. We then discuss possible physical 
improvements  of the friction model which might open the possibility of 
such dynamics, among which slip weakening of the friction 
coefficient, and stress the interest of developing systematic 
experimental investigations of this question.

\end{abstract}

\pacs{ 46.35.+z, 46.50.+a, 62.20.Mk, 81.40.Pq, 83.50.Fc}

\section{Introduction}
A few recent qualitative observations \cite{Rubio}, \cite{private}
 on the frictional motion of 
sheared gels sliding along smooth glass surfaces point towards the 
existence of inhomogeneous modes of frictional sliding. Namely, in 
some limited range of values of -small- shearing rates, sliding 
seems to occur via propagation of a quasi-periodic pattern of sliding 
zones of finite width, separated by non moving regions, where the 
interface sticks. These ``slip pulses'' drift at velocities 
$c\simeq mm/sec$, while the remote average (pulling) velocity lies in 
the $1-10 \mu m/sec$ range. Their width is typically tens of 
micrometers. Analogous observations have been made by 
Brune et al \cite{Brune} on a sliding rubber foam, and by Mouwakeh et 
al. \cite{Mouwakeh} on the elastomer polyurethane.

The topology of such sliding modes is reminiscent of that of 
Schallamach waves\cite{Schallamach}, which have been documented
\cite{Barquins} in the case of some 
very compliant transparent rubbers sliding on smooth glass. They 
consist of quasi periodic zones, of width typically $l\simeq 
100\mu$m, with space periods roughly $\sim 10l$ \cite{Olivier}, where the rubber 
buckles, so that the two surfaces get separated by a distance 
comparable with $l$. These separation waves have drift velocities 
$\sim$ mm/sec, for remote velocities $\sim \mu$m/sec. 

However, the
slip pulses in gels do not seem to be associated with any interface 
separation. In this respect, they are more comparable with the 
so-called ``self-healing slip pulses'', on which the attention of 
mechanicians has been focussing recently \cite{Ranjith}, following the suggestion by 
Heaton \cite{Heaton} that some major seismic events may have 
occurred, not by quasi 
simultaneous sliding of the whole rupture zone, but via fast 
propagation of localized sliding zones of small extent.

These observations all point towards a common question about the 
nature of frictional sliding, which can be schematized as follows. 
Consider two very thick blocks of solid materials with dissimilar 
elastic properties, in frictional contact along a planar interface of 
infinite lateral extent (Figure 1). This system bears a remote homogeneous 
compressive stress $\tau^{*}_{22}$, normal to the interface. Assume 
that, under the remote shear stress $\tau^{*}_{12}$, the upper block 
(I) slides towards $x_{1} > 0$ at a remote point velocity $v_{0}$ with 
respect to the lower one (II). Such motion can of course occur in a 
homogeneous mode, where stresses are uniform. Along the (homogeneous) 
interface, the friction law, which we assume to obey the 
Amontons-Coulomb proportionality between shear and normal loads, 
imposes that
\begin{equation}
\label{eq:1}
\tau^{*}_{12} = - f_{d}(v_{0})\tau^{*}_{22}
\end{equation}
If such is the case, given $\tau^{*}_{22}$ and $v_{0}$, the remote shear 
sliding stress is fixed. In the Coulomb approximation, where fine 
variations of the dynamic friction coefficient are ignored, $f_{d}$ 
reduces to a constant.

The question then arises of whether or not this homogeneous sliding 
mode is stable  with respect to small non homogeneous perturbations of 
the stress and strain fields localized in the surface region. In other 
words, do deformation waves exist along a sliding frictional 
interface? If so, are they damped, or amplified, or neutral? This question 
has been studied extensively, for dissimilar linear elastic 
materials, with Coulomb friction, by several authors, in particular 
Weertmann \cite{Weertmann}, Adams \cite{Adams}, and Martins et al 
\cite{Martins}, whose results are synthetized in 
a recent article by Ranjith and Rice \cite{Ranjith}. They find that, when 
$\mu_{d}\not=0$ and when such interface waves exist \cite{existence}, 
the corresponding sliding velocity field along the interface has, for 
a mode of wavelength $k$, the form :
\begin{equation}
\label{eq:2}
v\left(x_{1}\right) = 
v_{k}\exp\left[ik\left(x_{1}-ct\right)+a\mid k\mid t\right]
\end{equation}
Given the elastic moduli, the drift velocity $c$ and amplification 
coefficient $a$ are real positive constants. That is, waves drifting 
along (resp. against) the direction of $v_{0}$ are amplified (resp. 
damped). Homogeneous sliding is thus linearly unstable against 
perturbations of all wavelengths.

These results are derived under the assumption that the interface is
sliding everywhere. As amplification proceeds, the sliding velocity 
necessarily vanishes at some points. This suggests that sliding might 
occur via a periodic set of self healing slip pulses, separated by 
stick regions. A family of such pulses has been built by Adams 
\cite{Adams} for 
dissimilar elastic solids. Their drift velocity $c$, which depends on 
the values of elastic moduli, is, roughly speaking, on the order of a 
sound velocity. So, their dynamics is controlled by inertia. However, 
such self-sustaining (stationary) dynamical patterns are singular in 
the following sense. Since perturbations of all wavelengths, however 
small, are amplified, any initially localized perturbation gives rise 
to diverging oscillations at arbitrary small time : Adams's pulses 
have zero measure attractors. This so-called 
``ill-posedness'' most likely signals that the Coulomb friction law 
misses some of the physical processes which control the fast dynamics 
of fracture at frictional interfaces between elastic materials, i.e. 
their high frequency response - a problem which is currently under 
study \cite{Ranjith}.

Slip pulses in gels or rubbers, on the contrary, are slow 
dynamical objects whose 
velocities, comparable with those of Schallamach waves, are much too 
low for inertia to be relevant. Their dynamics is certainly controlled 
by the dissipation associated with the viscoelasticity of these 
materials.

We therefore concentrate, in this article, on the following question. 
Let block (I) be an incompressible linear viscoelastic material with, 
for simplicity, a single viscous relaxation time. It slides slowly on 
a smooth non-deformable material, and interface 
friction obeys a simple local Coulomb law. Under such conditions, are 
non inertial periodic slip pulses, stationary in the drifting frame, a 
possible mode of motion?

In Section II we formulate the corresponding mathematical problem, and 
derive the form of its analytical solutions. We show in Section III 
that none of these is compatible with the stick conditions to be 
satisfied in the non moving parts of the interface. Hence, in this as 
well as in the inertial regime, a Coulomb law with a constant 
dynamical friction coefficient is incompatible with the existence of 
such modes of motion. We discuss in Section IV possible physical 
tracks towards improvements of the simple Coulomb model, which might 
be relevant to the problem of inhomogeneous sliding, and stress the 
interest of corresponding experimental investigations.

\section{General Formulation}
We follow closely the approach of Adams\cite{Adams} and of Comninou 
and Dundurs \cite{CD} restricted to the 
case where (see Figure 1) block (II) $\left(x_{2}<0\right)$ is non 
deformable. Block (I) is submitted to the uniform remote stresses 
$\tau^{*}_{22}<0$ and $\tau^{*}_{12}$. It is infinitely extended 
along $x_{1}$, and made of an incompressible material, with a linear 
viscoelastic shear response described by :
\begin{equation}
\label{eq:3}
\tau_{12}\left(t\right) = \int 
\limits_{-\infty}^{t}dt'\,\mu\left(t-t'\right) 
\dot u_{12}\left(t'\right)
\end{equation}
$\tau_{12}$ and $u_{12}$ are the shear stress and deformation, 
confined to the $\left(x_{1}, x_{2}\right)$ plane, and we model the 
time dependent shear modulus as a single time Kelvin one, namely :
\begin{equation}
\label{eq:4}
\mu\left(t\right) = \mu_{\infty}+\left(\mu_{0}-\mu_{\infty}\right) 
e^{-t/\tau}
\end{equation}
We will moreover assume that the relaxed modulus $\mu_{\infty}$ is 
much smaller than the short time one, $\mu_{0}$. To fix ideas, for 
compliant rubbers, values of $\mu_{\infty}/\mu_{0} \lesssim 
10^{-3}$ are typical.

We want to study dynamic patterns, where (I) slides towards $x_{1} > 
0$ with the uniform remote velocity $v_{0}$, with space period 
$\lambda =2\pi/k$, and drift velocity $c$ in the frame of 
(II). The corresponding form of the displacements $u_{1},\,u_{2}$ 
reads :
\begin{equation}
\label{eq:5}
u_{1} = v_{0}t\,+\,\sum_{m\geq 
1}D_{m1}\left(x_{2}\right)\,e^{imk\left(x_{1}-ct\right)}
\end{equation}
\begin{equation}
\label{eq:6}
u_{2}= D_{02}
\,+\,\sum_{m\geq 1}D_{m2}\left(x_{2}\right)\,e^{imk\left(x_{1}-ct\right)}
\end{equation}
Solving the wave propagation equation together with the condition of 
non separation at the interface : 
$u_{2}\left(x_{1},\,x_{2}=0,t\right) = 0$, one obtains 
straightforwardly (see Appendix A), in the incompressible limit 
(Poisson coefficient $\nu = 1/2$) for the interfacial sliding 
velocity $v_{s}\left(\eta\right) = \partial u_{1}/{\partial 
t}\mid_{x_{2}=0}$ :
\begin{equation}
\label{eq:7}
v_{s}\left(\eta\right) = v_{0}\,+\,c Re\sum_{m\geq 1}{B_{m}
e^{im\eta}}
\end{equation}
where  $\eta = k\left(x_{1}-ct\right)$
while the interfacial shear and vertical stresses read :
\begin{equation}
\label{eq:8}
\tau_{12}\left(\eta\right) = \tau^{*}_{12} + Re\,\sum_{m\geq 1}
-iB_{m}\mu_{m}\left(1+\sigma_{m}\right) e^{im\eta}
\end{equation}
\begin{equation}
\label{eq:9}
\tau_{22}\left(\eta\right) = \tau^{*}_{22} + Re\,\sum_{m\geq 1}
B_{m}\mu_{m}\left(1-\sigma_{m}\right) e^{im\eta}
\end{equation}
with
\begin{equation}
\label{eq:10}
\sigma_{m} = \left[1-{{\rho 
c^2}\over{\mu_{m}}}\right]^{1/2}\,\,\,\,\,\,\,\,\,Re\sigma_{m}>0
\end{equation}
We restrict our attention to the very slow modes observed 
experimentally, for which, whatever the frequency 
$\left(mck\right)$,  $\rho c^2 << \mu_{m}$, so that
\begin{equation}
\label{eq:11}
\sigma_{m}\simeq {1 - {\frac{\rho c^2}{2\mu_{m}}}}
\end{equation}
$\rho$ is the mass density of material (I), $\mu_{m}$ 
its (complex) elastic modulus at frequency $\left(mck\right)$
\begin{equation}
\label{eq:12}
\mu_{m} = \hat{\mu}\left(mck\right) = \left[-i\omega \int 
_{0}^{\infty} dt\,\mu\left(t\right)\,e^{i\omega t}\right]_{\omega = 
mck}
\end{equation}

The unknown coefficients $B_{m}$ must be determined from the second 
boundary condition along the interface, which we want to describe a 
set of slipping regions of length $2l$ separated by sticking ones.

We assume friction to be described by a simple local Coulomb law, with 
a constant dynamic friction coefficient $f$ equal to the static one, 
that is :

(i) Slip regions : $-\alpha + 2p\pi < \eta < \alpha + 2p\pi $
\begin{equation}
\label{eq:13}
\tau_{s}\left(\eta\right) = \tau_{12}\left(\eta\right) + f 
\tau_{22}\left(\eta\right) = 0\,\,\,\,\,\,\,\,\,\,\,\,v_{s}>0
\end{equation}
(ii) Stick regions :  $-\alpha + 2p\pi < \eta < -\alpha + 
2\left(p+1\right)\pi$
\begin{equation}
\label{eq:14}
v_{s} = 0\,\,\,\,\,\,\,\,\,\,\,\,\,f\tau_{22} < \tau_{12} < -f\tau_{22}
\end{equation}

This description of interface friction calls for a few comments. 
Indeed, it assumes tacitly - as is common in contact mechanics 
\cite{KLJ} - that one can legitimately define a local 
and space-independent friction 
coefficient. Since solid friction results from the average effect of 
disipative flips of bistable pinned elastic units, this can be true 
only on a scale much larger than (i) the size $b$ of the basic unit, 
and (ii) the scale $L$ of interface inhomogeneities. The detailed 
analysis \cite{Berthoud}  of the Rice-Ruina phenomenological law 
of dynamic friction \cite{Rice-Ruina} 
has shown that $b$ is of nanometric order. So, our assumption is 
justified for interfaces with homogeneous intimate contact. Such is 
indeed the case for the gels or very compliant rubbers which we have 
in mind here, as long as elastic deformations vary on scales much 
larger than nanometers -- which sets a lower limit on the size of 
Dugdale-Barenblatt-like fracture head regions.

Note, however, that the situation is different when dealing with 
multicontact Greenwood-like interfaces \cite{Greenwood}. 
These prevail with stiff 
materials, such as metals, glasses or rocks, which are not polished
 down to nanometric 
roughness. Then, the small scale cutoff is provided by the average 
distance between contacting asperities, commonly lying in the $100 
\mu$m/sec range. This, in our opinion, should be kept in mind when 
attempting, for such interfaces, to regularize the above mentioned 
ill-posedness problem, since, on space scales $\lesssim L$ pinning 
strength fluctuations become non negligible.

Taking condition (\ref{eq:11}) into account we set, following 
Comninou and Dundurs \cite{CD}, for the periodic function 
$v_{s}(\eta)$, in $-\pi \leq \eta \leq \pi$ :
\begin{equation}
\label{eq:15}
v_{s}(\eta) = 0\,\,\,\,\,\,\,\,\,\,\alpha < \mid \eta\mid < \pi\,\,\,\,
\,\,\,\,\,(stick)
\end{equation}
\begin{equation}
\label{eq:16}
v_{s}(\eta) = v(\eta)\,\,\,\,\,\,\,\,\,-\alpha < \eta < \alpha\,\,\,\,
\,\,\,\,\,(slip)
\end{equation}
Hence, from equation (\ref{eq:6}) :
\begin{equation}
\label{eq:17}
B_{m} = {{1}\over{\pi c}}\int_{-\alpha}^{\alpha}d\xi \,v(\xi 
)e^{-im\xi}\,\,\,\,\,\,\,\,\,\,\,\,(m\geq 1)
\end{equation}
\begin{equation}
\label{eq:18}
v_{0} = {{1}\over{2\pi}}\int_{-\alpha}^{\alpha}d\xi\,v(\xi)
\end{equation}
Using equations (\ref{eq:15}), (\ref{eq:16}) together with 
(\ref{eq:8}), (\ref{eq:9}), and with the help of the relation
\begin{equation}
\label{eq:19}
\sum_{m\geq 1} e^{imx} = -{{1}\over{2}} + \sum_{n = - \infty}^{\infty} 
\delta(x-2n\pi)\,+\,{{i}\over{2}} PV\left[cotg{\frac{x}{2}}\right]
\end{equation}
(where $PV$ designates the Cauchy principal value), one finally gets, 
for the interfacial``sliding stress'' $\tau_{s}$ 
\begin{equation}
\label{eq:20}
\tau_{s}(\eta) = \tau_{s}^{*} + {{f\rho c^{2}}\over{2}}[V(\eta) - 
V_{0}] -{{1}\over{\pi}} \int_{-\alpha}^{\alpha}d\xi 
V(\xi)\,\intpv_{0}^{\infty}{ dt \mu(t) {{d}\over{dt}}\left[cotg{{\eta - \xi 
+ckt}\over{2}}\right]}
\end{equation}
where we have set $V(\eta) = {{v(\eta)}/{c}}$, and $\intpv f\,dx 
\equiv PV\int f\, dx$.

Integrating by parts the last term in the r.h.s. of Eq.(\ref{eq:16}), 
the condition Eq. (\ref{eq:10}) for frictional sliding within the slip 
pulses provides us with the integral equation to be satisfied by the 
interfacial reduced velocity field in $(-\alpha < \eta < \alpha)$, namely :
\begin{equation}
\label{eq:21}
\tau_{s}^{*}\,+\,{{f\rho 
c^{2}}\over{2}}\left[V(\eta)-V_{0}\right]\,+\,{{\mu_{0}}\over{\pi}}\intpv_
{-\alpha}^{\alpha}
d\xi\,V(\xi) cotg{{\eta - 
\xi}\over{2}}\,+\,{{1}\over{\pi}}\int_{-\alpha}^
{\alpha}d\xi\,V(\xi)\,\intpv_{0}^{\infty}dt\,{{d\mu}\over{dt}}
\left[cotg{{\eta - \xi 
+ckt}\over{2}}\right]\,=\,0
\end{equation}
with :
\begin{equation}
\label{eq:22}
\int_{-\alpha}^{\alpha}d\xi V(\xi) = 2\pi V_{0}
\end{equation}

Once Eqs.(\ref{eq:17}) are solved for $V(\eta)$, interfacial stresses 
in the stick regions should be calculated from Eq.(\ref{eq:16}), 
Eq.(\ref{eq:11}) then providing the final condition for slip pulses 
to exist.

Expression (\ref{eq:21}) separates explicitly the instantaneous 
elastic shear effects ($3^{rd}$ term) from the contribution of viscoelastic 
relaxation ($4^{th}$ term). The second term, which derives from the 
perturbation of the normal stress $\tau_{22}$, is, for our very slow 
pulses, smaller than the integral ones by a factor $(c/c_{s})^2$, 
where $c_{s}$ is some sound velocity. We will therefore neglect it 
from now on.

The cotg form of the elastic kernels results from imposing space 
periodicity to the patterns. Eq.(\ref{eq:21}) can be rewritten in a 
form more standard in fracture mechanics by setting, in ($-\pi < 
\eta < \pi$) :
\begin{equation}
\label{eq:23}
u = tg{\frac{\eta}{2}}\,\,\,\,\,\,y = 
tg{\frac{\xi}{2}}\,\,\,\,\,\,\,\,\,\,\Phi (u) 
={\frac{V(u)}{1+u^2}}\,\,\,\,\,\,a = tg{{\alpha}\over{2}} = 
tg{{kl}\over{2}}
\end{equation}
Some elementary algebra then leads, in ($-a < u < a$), to :
\begin{equation}
\label{eq:24}
{{\tau_{s}}\over{1+u^{2}}} = {{2\mu_{0}}\over{\pi}}\intpv _{-a}^{a} 
dy\,{\frac{\Phi(y)}{u-y}}\,+\,\int_{-a}^{a}dy\Phi(y) 
k(u,y)\,-\,{{2V_{0}u-\tau_{s}^{*}}\over{1+u^{2}}}\,=\,0
\end{equation}
with :
\begin{equation}
\label{eq:25}
\int_{-a}^{a}dy\Phi(y)\, =\,\pi V_{0}
\end{equation}
and we have set :
\begin{equation}
\label{eq:26}
k(u,y) = 
\frac{2}{\pi}\intpv_{0}^{\infty}dt\mu'(t)\frac{1+ytg(\frac{ckt}{2})}
{u-y+(1+uy)tg\left(\frac{ckt}{2}\right)}
\end{equation}

The singular integral equation (\ref{eq:24}) belongs to a class 
which was studied extensively by Mushkelishvili \cite{Mush}. In his 
terminology, the first two terms on the l.h.s. constitute the 
``dominant'' part. The viscoelastic kernel 
$k(u,y)$ satisfies the regularity condition : $lim_{u\rightarrow 
y}[(u-y)k(u,y)] = 0$. This entails that its plays no role in the 
strength of the singularities of the solutions -- i.e., as is 
intuitively reasonable, these are ruled by the instantaneous elastic 
response of the deformable medium.

Following reference \cite{Mush}, there are four families of solutions 
of equation (\ref{eq:24}), each of which is associated with one of 
the basic functions, characteristic of the dominant part :
\begin{equation}
\label{eq:27}
Z_{\epsilon_{t},\epsilon_{h}}(y) = 
(y+a)^{\epsilon_{t}/2}(a-y)^{\epsilon_{h}/2}
\end{equation}
where the indices $(\epsilon_{t,h}) = (+1,-1)$ control the convergent 
or divergent behavior of $Z$ at the tail and head edges of the slip 
zone.

One can then transform, for each family, Eq.(\ref{eq:24}) into an 
equivalent non-singular Fredholm integral equation. It moreover turns 
out that, when we specialize to the single relaxation time model 
for $\mu(t)$ (Eq.(\ref{eq:4})), analytical expressions for the 
solutions of these equations can be obtained explicitly, thus allowing 
us to draw explicit conclusions about their existence.

In view of the heaviness of the (otherwise straightforward) algebra 
involved, we will exemplify the method in full detail only for one of 
the families, namely the $(+-)$ one.
\section{The four families of solutions : 
$V$-fields and existence conditions}
\subsection{The $(+-)$ family}
The corresponding basic function
\begin{equation}
\label{eq:28}
Z_{+-}(y) = \sqrt {\frac {(y+a)}{(a-y)}}
\end {equation}
The implementation of Mushkelishvili's method is performed in 
Appendix B. For $\mu(t)$ as specified by Eq.(\ref{eq:4}), the non 
singular equation equivalent to Eq.(\ref{eq:24}) reads :
\begin{equation}
\label{eq:29}
\Phi_{+-}(u) = 
H_{+-}(u)\,+\,{{\Delta\mu}\over{\mu_{0}}}\,{{2exp(2tan^{-1}{u}/ck\tau)}
\over{ck\tau 
(1+u^2)}}\,\int_{u}^{a}dz\,\Phi_{+-}(z)\,\exp{\left(-2tan^{-1}{z}/ck\tau\right)}
\end{equation}
with :
\begin{equation}
\label{eq:30}
H_{+-}(u)\, 
=\,-\frac{Z_{+-}(u)}{2\mu_{0}}\,\frac{C^{*}u+D^{*}}{1+u^2}\,+\,\frac{\Delta\mu}
{\mu_{0}}
\,\frac{W_{+-}}{ck\tau}\,[\frac{Z_{+-}(u)}{\pi}G(u)+\frac{\beta}{1+u^2}
exp(2tan^{-1}u/ck\tau)]
\end{equation}
\begin{equation}
\label{eq:31}
W_{+-} = 2\int_{-a}^{a}dz\Phi(z) 
exp\left(-\frac{2tan^{-1}z}{ck\tau}\right)
\end{equation}
\begin{equation}
\label{eq:32}
G(u) = 
-\beta\int_{a}^{\infty}d\PsiÊ\frac{e^{2tan^{-1}\Psi/ck\tau}}{\left(u-\Psi\right)
\left(1+\Psi^2\right)}
\sqrt{\frac{\Psi-a}{\Psi+a}}\,-\,(1+\beta)
\int_{-\infty}^{-a}d\PsiÊ\frac{e^{2tan^{-1}\Psi/ck\tau}}{(u-\Psi)(1+\Psi^2)}
\sqrt{\frac{-\Psi+a}{-\Psi-a}}
\end{equation}
\begin{equation}
\label{eq:33}
\beta =(e^{2\pi/ck\tau}-1)^{-1}\,\,\,\,\,\,\,\,\,\,\,\Delta\mu = \mu_{0} 
-\mu_{\infty}
\end{equation}
\begin{equation}
\label{eq:34}
C^{*} = 2V_{0}\mu_{\infty}cos{\frac{\alpha}{2}}  - \tau_{s}^{*}sin{\frac{\alpha}{2}}
\end{equation}
\begin{equation}
\label{eq:35}
D^{*} =-2V_{0}\mu_{\infty}sin{\frac{\alpha}{2}}  - 
\tau_{s}^{*}cos{\frac{\alpha}{2}}
\end{equation}
Expression(\ref{eq:32}) for $G(u)$ is valid for solutions whose 
slip zone length $2l$ satisfies the condition $2l < \lambda/2$. We 
assume this to hold, in accordance with experimental observations, 
which indicate values of $2l/\lambda << 1$.

Eq. (\ref{eq:29}) is a first order differential equation for the 
function $\int_{u}^{a}dz\Phi_{+-}(z)exp[-2tan^{-1}z/ck\tau]$, which is 
straightforwardly solved into :
\begin{equation}
\label{eq:36}
\Phi_{+-}(u) = 
H_{+-}(u)\,+\,{\frac{2}{ck\tau}}{\frac{\Delta\mu}{\mu_{0}}}{\frac{1}{1+u^2}}\,
\int_{u}^{a}dz\,H_{+-}(z)\exp{\left[{\frac{\mu_{\infty}}{\mu_{0}}}
{\frac{2(tan^{-1}{u}-
tan^{-1}{z})}{ck\tau}}\right]}
\end{equation}

This defines a family of slip velocity fields, each of which is 
labelled by the four dimensionless 
parameters $V_{0}=v_{0}/c,\tau_{s}^{*}/\mu_{0}, l/\lambda = a/2\pi, 
c\tau/\lambda$. Two of the physical parameters, $v_{0}$ and 
$\tau_{s}^{*}$, are '`external'' : in an experiment, one imposes in 
general an average sliding velocity -- hence $v_{0}$ is fixed -- and 
measures $\tau_{s}^{*}$. $l, c, \lambda$, are the internal parameters 
of the family. This defines a problem of dynamical selection, namely : 
if sliding patterns exist, are $l, c, \lambda$, and hence $\tau_{s}$, 
uniquely defined when $v_{0}$ is fixed, or not? In order to clear up 
this important question, it is necessary to list the relations between 
them -- or, alternately, the conditions to be satisfied by $\Phi_{+-}$ 
as given by Eq.(\ref{eq:36}). These are :

(i) Two consistency conditions, expressing that the remote velocity 
and stresses are simply the $k = 0$ components of the corresponding 
fields. This is expressed by relation (\ref{eq:25}) and by an 
analogous equation for $\tau_{s}$ :
\begin{equation}
\label{eq:37}
\tau_{s}^{*} = 
{{1}\over{\pi}}\int_{-\infty}^{+\infty}du\,{{\tau_{s}(u)}\over{1+u^2}}
\end{equation}
where $\tau_{s}(u)$ is related to $\Phi_{+-}(u)$ by the first of 
equalities (\ref{eq:24}).

(ii)The interfacial stress field must also satisfy the stick 
inequality Eq.(\ref{eq:14}). One easily checks, with the help of 
Eq.(\ref{eq:24}), that a divergence of $\Phi$ at an edge, $u = \pm 
a$, of the slip zone results in a diverging $\tau_{s}$ at the 
corresponding stick zone edge, and therefore in the violation of the 
stick condition. $Z_{+-}$ (Eq.(\ref{eq:28})) diverges at the slip 
head $u = a$. For $u \rightarrow a$ :
\begin{equation}
\label{eq:38}
\Phi_{+-}(u) = \frac{Z_{+-}(u)}{2\mu_{0}}\,[-\frac{C*a+D*}
{1+a^2}+\frac{2\Delta\mu}{\pi}\,\frac{W_{+-}}{ck\tau}\,G(a)]\,+\,\Re(u)
\end{equation}
with lim$_{u\rightarrow a}\,\Re(u) = 0$.

A necessary condition for $\Phi_{+-}$ to be acceptable is that the 
coefficient of $Z_{+-}$ in Eq.(\ref{eq:38}) vanishes, i.e., using 
Eqs.(\ref{eq:29}-\ref{eq:35}) :
\begin{equation}
\label{eq:39}
\tau_{s}^{*}cos\frac{\alpha}{2} = 
-\frac{2}{\pi}\,\Delta\mu\,\frac{W_{+-}G(a)}{ck\tau}
\end{equation}
So, for solutions of the $(+-)$ class, the five pattern parameters 
are linked by three relations. That is, for a given $v_{0}$, this 
class of patterns, if they exist, form a one parameter family. We will 
comment further on this conclusion in section IV.

Let us now come back to the``regularization condition'' 
Eq.(\ref{eq:39}). From conditions (\ref{eq:15}) and (\ref{eq:16}), 
the interfacial sliding stress $\tau_{s}$ must be non positive 
everywhere. Hence, its $u$-average $\tau_{s}^{*}$, must be strictly 
negative.

On the other hand, the $v_{s}$, and thus the $\Phi$ field must, by 
Eq.(\ref{eq:13}), be positive everywhere in the slip zone. Then 
definition (\ref{eq:31}) entails that $W_{+-} > 0$. Finally, using 
Eq.(\ref{eq:32}), one gets:
\begin{equation}
\label{eq:40}
G(a) = 
-\,\int_{a}^{\infty}d\Psi\frac{sh[\frac{2}{ck\tau}(\frac{\pi}{2}-tan^{-1}
\Psi)]}{(1+\Psi^2)\sqrt{ \Psi^{2}-a^{2}}\, sh(\frac{\pi}{ck\tau})}\,<\,0
\end{equation}

Therefore, condition (\ref{eq:39}) can never be satisfied. No solution 
of type $(+-)$ exists. In other words, viscous relaxation and 
pulse-pulse interaction effects can never be sufficient to cancel the 
square-root singularity due to the instantaneous elastic response.
\subsection{The (-+) and (--) families}
For the $(-+)$ family, whose $Z$ function diverges at the slip zone 
tail only, the analysis parallels completely the above one, the 
$\Phi_{-+}$ fields obey a set of equations with exactly the same 
structure as that of Eqs.(\ref{eq:29}-\ref{eq:32}), differing only in the 
detailed algebraic expressions of $G(u)$, $C^{*}$, $D^{*}$. The 
regularization condition analogous to Eq.(\ref{eq:39}), now to be 
imposed at $u\rightarrow -a$, is again immediately shown to have no 
solutions.

$Z_{--}$ diverges at both slip edges, hence tow regularization 
conditions, and one shows similarly that the condition obtained from 
their difference cannot be satisfied. Note, however, that the 
counting argument tells us that, if $(--)$ patterns could exist, one 
only of the five parameters would be free, i.e. there could exist at 
most one dynamical pattern at a given sliding velocity.
\subsection{The (++) family}
As $Z_{++}$ vanishes at both slip zone edges, no regularity condition 
has to be imposed \cite{note}. (++) solutions, if any, form a two 
parameter family.. The analysis of Appendix B again leads to an 
expression of $\Phi_{++}$ with the same structure as Eqs. 
(\ref{eq:29}-\ref{eq:32}). One can then write explicitly the self consistency 
equation (\ref{eq:25}) for $V_{0}$. We will skip here the 
corresponding tedious but straightforward algebra, and only quote the 
final form of Eq.(\ref{eq:25}), which can be written as :
\begin{equation}
\label{eq:41}
V_{0}\left[1+O\left({{\mu_{\infty}}\over{\mu_{0}}}\right)\right] = 
{{\tau_{s}^{*}}\over{2\mu_{0}}}\,{{1}\over{ck\tau}}\,{(1-cos{ 
{{\alpha}\over{2}}})^2}{cos{{\alpha}\over{2}}}
\end{equation}
For the systems we are interested in, as already mentioned, 
$\mu_{\infty}/\mu_{0} << 1$. Then, again, under the stick restriction 
which imposes that $\tau_{s}^{*} < 0$, condition (\ref{eq:29}) cannot 
be fulfilled.
\section{Discussion}
The above analysis leads us to a strong statement, which seems to 
contradict existing qualitative observations. Namely, an interface 
with Coulomb friction between a viscoelastic and a non-deformable 
material cannot sustain slow sliding via a periodic set of alternating 
non inertial slip pulses and stick regions.

We believe that the reason for this contradiction must be traced to 
the fact that the Coulomb model of friction which we have assumed 
misses some physical elements which probably play a crucial role in 
the dynamics of patterns with fracture-like singularities. This may 
appear more clearly when one notices that this model, which describes 
the interface as infinitely rigid below the friction threshold, then, 
once this is reached, sliding under constant stress, is the exact 
$2D$ equivalent of the Hill model of bulk ideal plasticity \cite{Hill}, 
well known 
to generate numerous artefact instabilities due to its highly 
singular character.

Clearly, the main weakness of the Coulomb model lies in its 
overschematic description of the transition between stick and slip 
i.e., for our patterns, in the details of interface boundary 
conditions at the edges of a slip zone. In the case of the analogous 
mode-I problem -- the Griffith crack -- it is well known that 
discontinuous boundary conditions (vanishing normal stresses and 
displacements in, respectively, the cracked and uncracked regions of 
the crack plane) miss a major physical ingredient, namely the finite 
range of atomic decohesion and its associated energy cost. Taking this 
into account regularizes the stress field at the fracture head, by 
smearing its square root singularity over the Dugdale-Barenblatt 
cohesive zone \cite{LL} .

In analogy with this, and based upon the nature of the 
stress-strain characteristics 
of overconsolidated clays, Palmer and Rice \cite{PalmerRice} proposed 
a model for sliding along a concentrated slip surface in which the 
sliding shear stress is assumed to decrease with relative displacement 
as shown on Figure 2. This enabled them to analyze the ``mode-II 
fracture'' problem of shear band propagation in such materials.

Let us assume for the moment that we can modify our Coulomb model in 
a similar manner. That is, let us assume that the shear stress in the 
sliding state is given by :
\begin{equation}
\label{eq:42}
\tau_{12} = - f\tau_{22} + \delta\tau_{0}[\delta(x_{1},t)]
\end {equation}
where $\delta\tau_{0}$ is maximum for $\delta = 0$ and has a small 
range $\delta_{0} << 2l$. We define $\delta(x_{1},t)$ as the 
displacement at the interface point $x_{1}$ from its position when it 
was in the preceding stick zone, i.e. up till the head of the slip 
zone under consideration reached it. So :
\begin{equation}
\label{eq:43}
\delta(x_{1},t) = u_{1}(x_{1},t) -  u_{1}(x_{1},t-\frac{a-x_{1}}{c}) 
= \int_{\eta}^{\alpha}d\eta'\,V(\eta')\equiv \delta(\eta)
\end{equation}
In Eq.(\ref{eq:21}) for the velocity field, $\tau_{s}^{*}$ must now 
be substituted by $\tau_{s}^{*} - \delta\tau_{0}[\delta(\eta)]$.

In order to fix ideas, let us concentrate on $(+-)$ solutions. 
Repeating the analysis of Appendix B leads again to expression
(\ref{eq:36}), with :
\begin{equation}
\label{eq:44}
H_{+-}(u)\,\rightarrow\,H_{+-}(u) - \frac{Z_{+-}(u)}{2\mu_{0}} I(a)
\end{equation}
\begin{equation}
\label{eq:45}
I(a) = 
-\frac{1}{\pi}\intpv_{-a}^{a}dy\,\frac{\delta\tau_{0}[\delta(y)]}{(1+y^2)
(a-y)Z_{+-}(y)}
\end{equation}
The regularity condition then becomes :
\begin{equation}
\label{eq:46}
\tau_{s}^{*}\,cos{\frac{\alpha}{2}}\,+\,\frac{2\Delta\mu}{\pi} 
\frac{W_{+-}G(a)}{ck\tau} = \frac{I(a)}{\pi}
\end{equation}
The l.h.s. of Eq.(\ref{eq:46}) must, as shown above, be negative. 
From Eq.(\ref{eq:45}), $I(a) < 0$. So, the introduction of a ``cost 
for incipient sliding'' via $\delta\tau_{0}$ is sufficient to lift 
the incompatibility which we found to hold for the Coulomb friction 
model. Clearly, the same formal result applies for the (-+) and (--) 
classes. That is, the localized incremental stress $\delta\tau_{0}$ 
plays a role comparable to that of the cohesive stress in mode-I 
fracture, namely it smoothes out the stress singularity by spreading 
it over a zone of incipient sliding of small but finite extent.

Once this formal remark has been made, one should however come back to 
the possible physical interpretation of such a modification of the 
friction model. A decrease in the frictional stress with the slip 
distance is likely to be associated with a change upon sliding of the 
internal structure of the nanometer-thick adhesive interfacial junction. 
Moreover, in order for the peaked structure of $\tau_{12}(\delta)$ to 
reproduce itself at each successive slip zone head, the structure of 
the junction must relax non negligibly on the duration 
$\Delta\tau_{st} = (\lambda-2l)/c$ of a stick (typically, 
$\Delta\tau_{st}$ lies in the range of seconds).

Such a scenario is plausible for junctions composed of long molecules 
-- either because a molecular layer of lubricant is present or because 
the junction is formed by molecular tails from the sliding material 
itself. Then, sliding is likely to give rise to a slip weakening of 
friction associated with molecular elongation and restrengthening by
structural relaxation during stick. These are precisely the physical 
ingredients invoked to explain the hysteretic frictional dynamics 
observed in a number of boundary lubrication experiments \cite{Jacob}, 
\cite{Carlson}, \cite{Josephson}   .

However, inclusion of slip-weakening of dynamical friction is not the 
only possible improvement on the Coulomb model susceptible to allow 
for slip pulses. Indeed, a series of recent works by Langer et al 
\cite{Langer} \cite{Lobk} on the viscoplasticity of amorphous solids point very 
convincingly towards the crucial importance of a realistic 
description -- of the rate and state type -- of the gradual 
cross-over of the mechanical shear response from mainly elastic to 
mainly dissipative. As already pointed out, solid friction along a 
continuous interface is nothing but $2D$ interfacial viscoplasticity, 
to which the bulk analysis should be transposable. Hence the need for 
the elaboration of a phenomenolgy which can bridge realistically between 
static and dynamic solid friction. Work in that direction, based upon 
experimental studies of dynamic interfacial shear response, is 
presently in progress.

This discussion naturally leads us to emphasize the need for the 
development of systematic experimental studies of interfacial slip 
pulses, and the interest which they present. The main questions to be 
elucidated are :

(i) the precise conditions for frictional sliding to occur in this mode. 
This includes systematic characterization of the bulk viscoelasticity 
of systems which do exhibit this behavior, and qualification of the 
relevant range of driving velocities $v_{0}$.

(ii) the $v_{0}$-dependence of the apparent friction coefficient 
$\tau_{12}^{*}/\mid \tau_{22}^{*}\mid$, and the question of pattern 
selection. Namely, is the slip pattern unique for a given $v_{0}$, 
or, for example, does it depend on the lateral size of the sliding 
block? In other words, does injection at the back free edge of the 
slider play a crucial role in the selection of the pattern 
wavelength $\lambda$, or not?

Further elucidation of these questions would also be of value to shed 
further light on the still largely open question of the physics of 
shear interfacial fracture.
  
\acknowledgments
It is a pleasure to thank J.R. Rice for a number of illuminating 
discussions about this and related subjects. I am indebted to 
T.Baumberger for drawing my attention to this question, and for 
permanent exchange during the course of this work, and to O. Ronsin 
and B .Velicky for fruitful discussions.

\appendix
\section{}
We briefly sketch here the derivation of eqs.(\ref{eq:7}-\ref{eq:9}) 
for a purely elastic system.
Let $\lambda, \mu$ be its Lame coefficients, related to the Young 
modulus E and to the Poisson ratio $\nu$ by :
%
%
\begin{equation}
\label{eq:A.1}
\mu ={\frac{E}{2\left(1+\nu\right)}}\,\,\,\,\,\,\,\,\,\,
\lambda + \mu = {\frac{E}{2\left(1+\nu\right)\left(1-2\nu\right)}}
\end{equation}
%
%
The elastic displacement ${\bf u} = (u_{1},u_{2})$ obeys the Lame 
equation :
\begin{equation}
\label{eq:A.2}
\rho {\bf \ddot{u}} = \left(\lambda + \mu \right)\,{\nablabf}. 
div{\bf u} + \mu \Delta{\bf u}
\end{equation}
with $\rho$ the mass density. The stresses are given by :
\begin{equation}
\label{eq:A.3}
\frac{\tau_{ij}}{\mu} = \frac{\partial u_{i}}{\partial x_{k}} +
\frac {\partial u_{k}}{\partial x_{i}} +\left(\beta^{2} - 
2\right)\delta_{ik}\, div\bf u
\end{equation}
where :
\begin{equation}
\label{eq:A.4}
\beta^{2} =\frac{2\left(1-\nu\right)}{1-2\nu}
\end{equation}
One then sets :
\begin{equation}
\label{eq:A.5}
u_{i}\left(x_{1},x_{2},t\right) = u_{i}^{*}\left(x_{2}\right) + 
\sum_{m}U_{im}\left(x_{2}\right) e^{imk\left(x_{1}-ct\right)}
\end{equation}
with ${\bf u}^{*}$ the displacement field corresponding to uniform 
sliding under the homogenenous stresses $\tau^{*}$. Solving equation
(\ref{eq:A.2}) together with the condition of non separation at the 
interface $u_{2}\mid _{x_{2}=0} = 0$, one gets :
\begin{equation}
\label{eq:A.6}
u_{1}\left(x_{1},x_{2},t\right) - u_{1}^{*}\left(x_{2}\right) = Re \sum _{m\geq 1}
A_{m}\left[ - \frac{k^{2}}{s_{+}s_{-}} e^{-ms_{+}x_{2}}\, +\, 
e^{-ms_{-}x_{2}}\right]e^{imk\left(x_{1}-ct\right)}
\end{equation}
\begin{equation}
\label{eq:A.7}
u_{2}\left(x_{1},x_{2},t\right) - u_{2}^{*} = Re \sum _{m\geq 1}
A_{m} \frac{ik}{s_{-}}\left[ -  e^{-ms_{+}x_{2}}\, +\, 
e^{-ms_{-}x_{2}}\right]e^{imk\left(x_{1}-ct\right)}
\end{equation}
with :
\begin{equation}
\label{eq:A.8}
s_{+}^{2} = k^{2}\left(1-\frac{\rho c^{2}}{\lambda + 
2\mu}\right)\,\,\,\,\,\,\,\,\,s_{-}^{2} = k^{2}\left(1-\frac{\rho 
c^{2}}{\mu}\right)\,\,\,\,\,\,\,\,Re\,s_{\pm} > 0
\end{equation}
Then, with the help of Eq.(\ref{eq:A.3}), and in the incompressible 
limit $\lambda \rightarrow \infty$, one obtains the 
expression for the interface stresses and the interface sliding 
velocity :
\begin{equation}
\label{eq:A.9}
\tau_{12}\mid _{x_{2}=0} = \tau_{12}^{*} + \mu Re \sum_{m\geq 1}
mkA_{m}\left(\frac{k}{s_{-}}-\frac{s_{-}}{k}\right)
e^{imk\left(x_{1}-ct\right)}
\end{equation}
\begin{equation}
\label{eq:A.10}
\tau_{22}\mid_{x_{2}=0} = \tau_{22}^{*} + \mu Re \sum_{m\geq 1}
imkA_{m}\left[-2 + \frac{k}{s_{-}}+\frac{s_{-}}{k}\right]
e^{imk\left(x_{1}-ct\right)}
\end{equation}
\begin{equation}
\label{eq:A.11}
v_{s} =v_{0} - c Re \sum_{m\geq 
1}imkA_{m}\left[1-\frac{k}{s_{-}}\right]e^{imk\left(x_{1}-ct\right)}
\end{equation}
Setting :
\begin{equation}
\label{eq:A.12}
B_{m} =imk A_{m}\left(\frac{k}{s_{-}}-1\right)
\end{equation}
and substituting $s_{-}$ by $s_{m-} = k \sigma_{m}$ 
(Eq.(\ref{eq:A.3})) appropriate to the viscoelastic system directly 
yields expressions (\ref{eq:7})-(\ref{eq:11}).

\section{}
Following \cite{Mush}, the singular integral equation (\ref{eq:24}), 
valid in $-a <u <a$ :
\begin{equation}
\label{eq:B.1}
{{2\mu_{0}}\over{\pi}}\intpv _{-a}^{a} 
dy\,{\frac{\Phi(y)}{u-y}}\,+\,\int_{-a}^{a}dy\Phi(y) 
k(u,y)\,=\,F(u)
\end{equation}
where $k$ is given by Eq.(\ref{eq:26}) and :
\begin{equation}
\label{eq:B.2}
F\left(u\right) = \frac{2V_{0}u-\tau_{s}^{*}}{1+u^{2}}
\end{equation}
is equivalent, for the $(+-)$ family of solutions, to :
\begin{equation}
\label{eq:B.3}
\Phi\,+\,K\star k\star \Phi = K\star F
\end{equation}
with 
\begin{equation}
\label{eq:B.4}
\left[K\star f\right]\left(u\right) = -\,\frac{Z_{+-}\left(u\right)}{2\mu_{0}\pi}
\intpv _{-a}^{a} dy\, 
\frac{f\left(y\right)}{Z_{+-}\left(y\right)\left(u-y\right)}
\end{equation}
Integrating in the complex y-plane along the contour shown on Figure 
3, one finds :
\begin{equation}
\label{eq:B.5}
K\star F = -\frac{Z_{+-}\left(u\right)}{2\mu_{0}}\,\frac{C^{*}u+D^{*}}
{1+u^{2}}
\end{equation}
where $C^{*}, D^{*}$ are given by equations (\ref{eq:34}), (\ref{eq:35}).

On the other hand :
\begin{equation}
\label{eq:B.6}
\left(K\star k\star \Phi\right)\left(u\right) =
-\,\frac {Z_{+-}\left(u\right)}{2\mu_{0}\pi}\int_{-a}^{a}
dz \Phi\left(z\right) J_{\rightarrow}\left(y,z\right)
\end{equation}
\begin{equation}
\label{eq:B.7}
J_{\rightarrow}\left(y,z\right) = \frac{2}{\pi}\,\intpv_{-a}^{a}\frac
{dy}{Z_{+-}\left(y\right)\left(u-y\right)}\,\intpv_{0}^{\infty}
ds\,\frac{\mu'\left(s\right)}{y-\frac{z-T\left(s\right)}{1+zT\left(s\right)}}
\end{equation}
with :
\begin{equation}
\label{eq:B.8}
T\left(s\right) = \tan
 \frac{cks}{2}
\end{equation}
Once the order of the $y$ and $s$-integrals on the r.h.s. of 
eq.(\ref{eq:B.7}) has been interchanged,the $y$-integration can be 
performed explicitly. However, care must be exercised when performing 
this interchange, due to the presence of the two principal values. 
One uses the following identity, which results from the Poincare-Bertrand 
theorem\cite{Mush}:
\begin{eqnarray}
\label{eq:B.9}
PV\left(\frac{1}{x''-x'}\right) PV\left(\frac{1}{x''-x}\right)&=&
PV\left(\frac{1}{x-x'}\right)\left[PV\left(\frac{1}{x''-x}\right)-
PV\left(\frac{1}{x''-x'}\right)\right]\,\nonumber \\&&\quad +\,\pi^{2}\delta\left(
x''-x'\right) \delta\left(x''-x\right)
\end{eqnarray}
One thus obtains:
\begin{equation}
\label{eq:B.10}
J_{\rightarrow}\left(y,z\right)= 
\intpv_{0}^{\infty}ds\mu'\left(s\right)\intpv_{-a}^{a}\,\frac{dy}{Z_{+-}
\left(y\right)\left 
(u-y\right)\left(y-\frac{z-T\left(s\right)}{1+zT\left(s\right)}\right)}\,\,+\,\,Y
\end{equation}
\begin{equation}
\label{eq:B.11}
Y = 
-\pi^{2}\,\sum_{p=-\infty}^{\infty}\frac{\mu'\left(s_{p}
\left(z,u\right)\right)}{\mid \partial D/\partial s\mid_{s_{p}}\,Z_{+-}
\left(u\right)}\theta\left( s_{p}
\left(z,u\right)\right)
\end{equation}
where the $s_{p}$ are the zeros of $D = u-\frac{z-T\left(s\right)}
{1+zT\left(s\right)}$, i.e.:
\begin{equation}
\label{eq:B.12}
s_{p}\left(z,u\right) = 
\frac{2}{ck}\left[\phi\left(z,u\right)+p\pi\right]
\end{equation}
\begin{equation}
\label{eq:B.13}
\phi\left(z,u\right) = 
\tan^{-1}\left[\frac{z-u}{1+zu}\right]\,\,\,\,\,\,\,\,-\frac{\pi}{2}
<\phi\left(z,u\right)<\frac{\pi}{2}
\end{equation}

From this one gets finally :
\begin{eqnarray}
\label{eq:B.14}
\left(K\star k\star \Phi\right)\left(u\right) &=& 
\frac{2}{\mu_{0}ck\left(1+u^{2}\right)}\int_{-a}^{a}dz\Phi(\left(z\right)
\mu'\left[s_{0}\left(z,u\right)\right]\left[\theta\left(\phi\left(z,u\right)
\right)+\beta\right]\,\nonumber\\&&\quad -\,\frac{Z_{+-}\left(u\right)}{2\mu_{0}}\int_{-a}
^{a}dz\Phi\left(z\right)\intpv_{0}^{\infty}ds\frac{\mu'\left(s\right)
\theta\left(\psi^{2}-a^{2}\right)}{Z_{+-}^{out}\left(\psi\right)\left(
u-\psi\right)}
\end{eqnarray}
where $\beta$ is defined in eq.(\ref{eq:33}), and :
\begin{equation}
\label{eq:B.15}
Z_{+-}^{out}\left(\psi\right)=\theta\left(\psi 
-a\right)\sqrt{\frac{\psi+a}{\psi -a}}\,+\,\theta\left(-\psi 
-a\right)\sqrt{\frac{-\psi-a}{-\psi +a}}
\end{equation}

Straightforward integration then results in 
eqs.(\ref{eq:29}-\ref{eq:32}).

\begin{figure}
\caption{
\label{fig:1}
Schematic representation of the sliding system}
\end{figure}
\begin{figure}
\caption{
\label{fig:2}
Schematic representation of the slip-weakening friction law.}
\end{figure}
\begin{figure}
\caption{
\label{fig:3}
Contour for integrals of type $K\star F$.}
\end{figure}


\begin{references}
\bibitem{existence} The precise conditions for their existence are 
determined by the value of the friction coefficient and, as is the 
case for interface elastic waves at a cohesive interface, by the 
values of the ratios of elastic moduli.
\bibitem{note} In classical mode-I fracture problems such as that of 
the Griffith crack, such a solution has to satisfy an auxiliary 
condition which expresses the fact that the perturbation of the stress 
field vanishes at infinity. There is no such condition for our 
problem -- where it is replaced by condition (\ref{eq:11}), which must 
hold everywhere in the stick zones (note that $u\rightarrow\pm\infty$ 
corresponds to their midpoints)
\end{references}
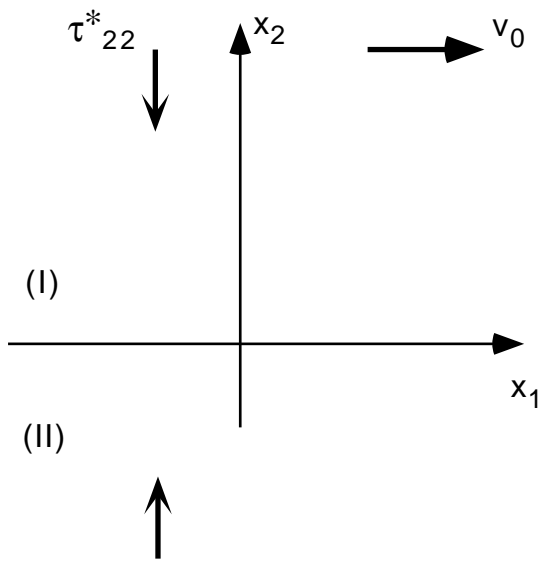
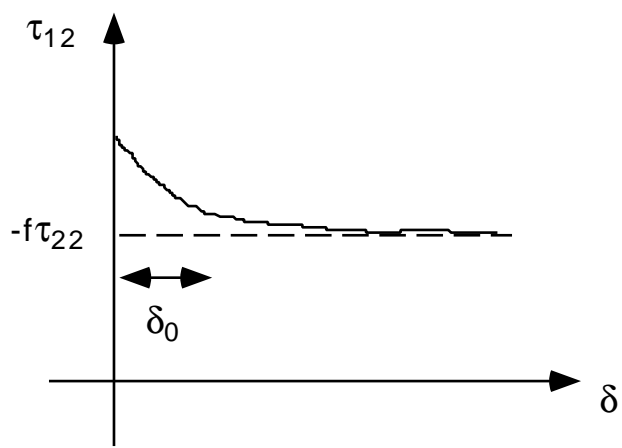
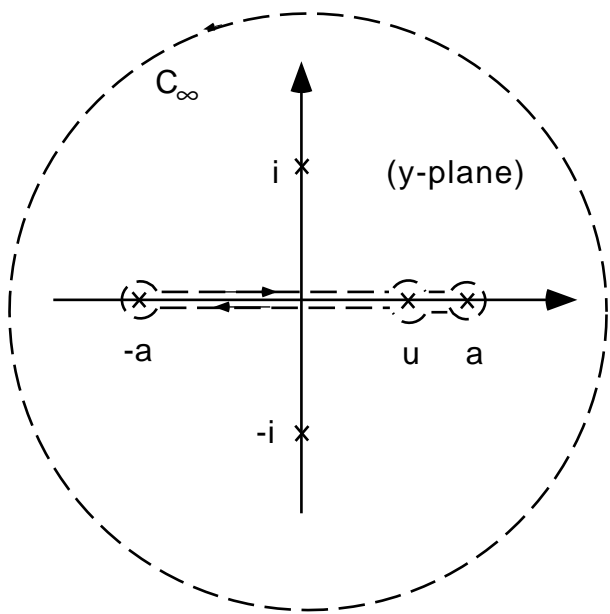
\end{document}